# From Single Helical relaxed states to helical equilibria

R. Paccagnella

Consorzio RFX
and  Istituto Gas Ionizzati del CNR
Corso Stati Uniti 4,
35127  Padova, Italy

**Abstract:**

This paper analyzes the relationship between the Single Helical relaxed states studied in [1] and the so called "helical ohmic equilibria", i.e. plasma states that are solutions of the helical Grad Shafranov equation and that satisfy the constraint imposed by the Ohm's law. The existence of marginally stable helical equilibria is clearly demonstrated, while the ohmic constraint is not exactly satisfied within the proposed first order perturbation theory. The model predictions are however satisfactorily compared with experimental trends.

**Introduction:**

In the past the discovery of states with dominant modes (i.e. dominant helicities) in various Reverse Field Pinch (RFP) devices has attracted considerable interest. In experiments these states are never perfectly single helicity, since subdominant modes are also present. This complicates the description of the system in particular if questions related to improvement of the plasma confinement properties and the related increase of the plasma global thermal content, in such quasi single helical states, are raised [2].

While experimentally pure single helicity states ( and therefore with a perfect helical symmetry) are hardly obtained, it is possible and instructive to study them in  RFP plasmas just from a  theoretical point of view.  This approach  can help in identifying features of the single helical plasmas that can then be compared with the experimental results and trends.

In particular, the purpose of this paper is to consider the helical relaxed states (SHR from now on) recently studied in [1] as an initial guess for the helical Grad-Shafranov equation (HGS) and to check, a posteriori, if the obtained helical states are satisfying the Ohm's law.

Clearly if such a state could be constructed it might be the closest to the experimentally observed quasi single helicity states.

The SHR are states of minimum magnetic energy under conservation of global helicity, toroidal flux and the helical invariant proposed in [3,4]. These states have been shown to predict successfully some of the experimental results obtained in RFPs [1].

These states are cylindrical symmetric, i.e. no information is obtained about the structure (eigenfunction) of the dominant mode, apart from its periodicity, which is an input to the model.

The coupled solution of SHR and HGS can instead add important information about the profiles of the helical perturbed magnetic fields in the plasma and about the associated magnetic island.

In order to solve the coupled system we propose for the HGS equation a perturbative solution. Therefore the SHR are given as initial state to the HGS solver and iteratively modified magnetic fields, including the non-axisymmetric perturbation, can be obtained.

Once the HGS solution has been found its compatibility with the Ohm's law, in order to test the existence of a so called "ohmic equilibrium state", is checked. It was found that the satisfaction of the constraint is only approximately achieved within our first order perturbative approach..

The paper is organized as follows:

section 1 briefly summarizes the content of the relaxation theory and some of the already obtained results, section 2 is describing our perturbative approach in solving the HGS equation, section 3 is giving the main numerical results and finally a discussion and some conclusions are presented.

**1. Single helical relaxation**

The relaxation theory proposed in [3,4] assumes in the plasma a dominant mode with a given helicity.

It corresponds to a helical perturbation having a poloidal, m, and toroidal, n, mode numbers.

A minimization procedure for the energy subjected to the conservation of total helicity and of an invariant related to the dominant mode expressed as [4]:

$$K_d = \frac{1}{2} \int_V \chi^d \mathbf{A} \cdot \mathbf{B} \, dV$$

where the integral is over the whole plasma volume, $\mathbf{A}$ and $\mathbf{B}$ are respectively the magnetic potential and the magnetic field and $\chi$ is the helical flux function of the mode defined as:

$$\chi = q_S \Psi - \Phi$$

where $q_s$ is the mode pitch, $\Psi$ is the poloidal flux, while $\Phi$ is the toroidal flux, d is zero or a positive integer. The idea is that the invariant $K_1$ ( beside $K_o$ i.e. the total helicity ) is a well preserved quantity in a relaxation process dominated by a single mode. The exponent d is a free parameter that is introduced [4] in order to let the solutions to be independent of the magnetic field normalization..

The relaxed state is shown to satisfy the equation:

$$\mathbf{J} = \sum_d \frac{\lambda_d \, (d+2)}{2} \chi^d \mathbf{B} \qquad (1)$$

where d is a null or positive integer (as said d=0 corresponds to the conservation of the total helicity) and the $\lambda_d$'s are eigenvalues. (Note also that in Ref. [1] the summation symbol in Eq.(3) (equivalent to Eq.(1) ) was missed by mistake).

We are assuming in Eq.(1) either d=0,1 or d=0,2 and simplifying the search only to the lowest eigenvalue $\lambda_o$ (by selecting a simple algebraic dependence of $\lambda_1$ or $\lambda_2$ from $\lambda_o$ ). In [1] the theory was applied to an intermediate aspect ratio A ( A=R/a where R and a are respectively the major and minor radius of the reference torus). It has been shown that the choice d=2 can reproduce reasonably well the experimental trend at A=4 - 5. It has also been found that a cooperative use of the relaxation model described above (thereafter called SHR model) and of a two region partial relaxation model (named hereafter TR) (see [1] for details) has allowed predictions in good agreement with the experiments.

In this paper, as anticipated, the axi-symmetric parallel current profile $\lambda_{axi}$ (also denoted later as $\lambda_{0,0}$) deduced within the SHR model are used to initialize the profiles of the helical Grad-Shafranov equation as described in more details in the next section.

## 2. Perturbative solution of the HGS equation with ohmic constraint:

The HGS represents the magneto-static plasma equilibrium in presence of a helical symmetry. We consider the problem of the helical equilibrium in cylindrical geometry.

With this assumptions the magneto-static equation:

$$\mathbf{J} \times \mathbf{B} = \nabla p$$

where J is the current density, B the magnetic field and p pressure in the plasma region, can be re-written as:

$$L_h \chi = \frac{1}{f}\frac{\partial}{\partial r}\left(f\frac{\partial \chi}{\partial r}\right) + \frac{1}{rf}\frac{\partial^2 \chi}{\partial u^2} = (\beta - \lambda(\chi))\,g(\chi) - \frac{r}{f}\frac{\partial p}{\partial \chi} \qquad (2)$$

with $g(r,u) = r\vec{\sigma}\cdot\vec{B} = mB_z - krB_\theta$ being the helical magnetic field,

$\chi(r,u) = r\vec{\sigma}\cdot\vec{A} = mA_z - krA_\theta$ being the helical flux function,

and $f(r) = \dfrac{r}{m^2 + k^2 r^2}$ , $\beta = \dfrac{-2km}{m^2 + k^2 r^2}$ and $u = m\vartheta + kz$ is the helical angle

with m the poloidal number of the helix and the cylindrical mode k = -n/R with n the toroidal mode number and R the torus major radius, to account for the toroidal periodicity in the real systems. We mention that our analysis is limited to the poloidal mode number m=1, in agreement with the experimental observations in RFPs.

Our aim is to initialize Eq.(2) from the axi-symmetric profiles obtained as solution of the single helical relaxed states. Therefore a perturbative approach is the most appropriate.

It has been observed [5] that the marginal MHD Newcomb equation is obtained by linearizing the generalized Grad-Shafranov equation for helical equilibria. Using this fact, a perturbative scheme (see the block-diagram below ) is possible. Initially the axisymmetric SHR state is chosen and on the top of this state an helical equilibrium is built iterating the relevant Grad-Shafranov equation and the Newcomb equation.

A complication is related to the resonant modes. For these modes in fact the marginal MHD equation is singular near the resonance surface. This problem has been circumvented by a suitable smoothing of the solution of the ideal marginal MHD equation near the resonance. This procedure gives generally solutions with continuous derivatives only when a maximum of the $\lambda$ exist near the resonance of the mode. This fact can be understood if we consider the following equation:

$$\lambda_1 = \frac{-r b_r \frac{\partial \lambda_o}{\partial r}}{\frac{\partial \chi_o}{\partial r}} \quad with \quad \left.\frac{\partial \chi_o}{\partial r}\right|_{rs} = 0 \qquad (3)$$

which is derived from the div $\mathbf{J}$ =0 condition for force-free equilibria considering only terms up to the first helical harmonic. The symbol 'o' refers to the axi-symmetric component of the quantities.

From Eq.(3) it is clear that a continuous solution for the perturbed current can be obtained only when, for a finite $b_r$, the derivative of the axisymmetric $\lambda$ is zero where the mode is resonant.

However an interesting physical point arises: besides the continuous solutions of the helical Grad-Shafranov equation, also solutions with current sheets near the resonant surface may exist. A similar general consideration brought us earlier to build the TR solution discussed in [1].

To perturb the initial axi-symmetric equilibrium an arbitrary parameter, $\varepsilon$, should be chosen (see the flow chart) todetermine the amplitude of the radial magnetic field (and therefore the amplitude of the associated island). Since our approach is perturbative the value of $\varepsilon$ cannot be arbitrarily large. We have however verified that the scheme is reasonably working for values corresponding to relatively large magnetic islands, up to 30 % of the plasma minor radius.

## HGS solver flow chart

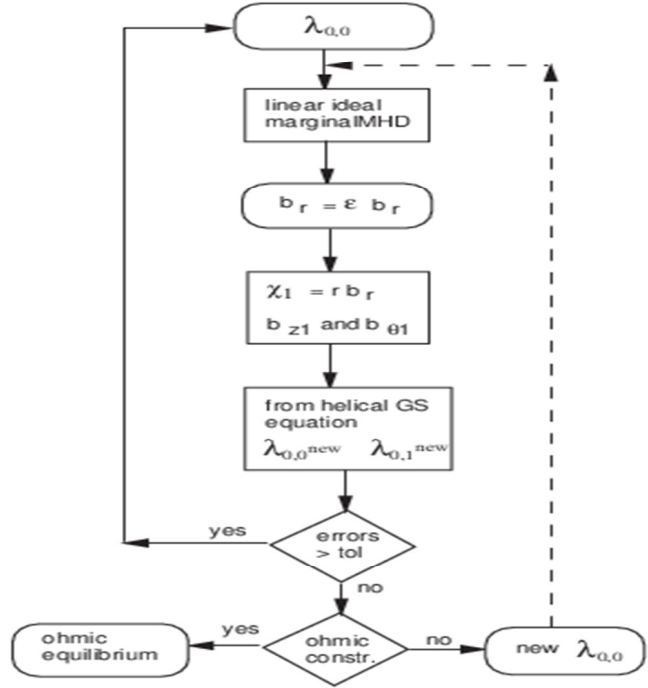

The flow-chart schematically describes our procedure. The initial axi-symmetric $\lambda_{o,o}$ profile is the axi-symmetric SHR solution, from it the Newcomb equations is solved [5] and the profile for the radial magnetic field is deduced. After multiplication by the free chosen $\varepsilon$ parameter the radial field amplitude is set. The first order in $\varepsilon$ perturbation of the helical flux function, $\chi_1$, is calculated together with the other two perturbed components of the magnetic field. Than from the HGS equation a new $\lambda_{o,o}$ and a new perturbed quantity $\lambda_{o,1}$, which defines. the first order helical correction to the parallel current, is calculated. This procedure is iterated till a certain specified convergence, i.e. till the point at which the i+1 iterated profiles differ less than a specified tolerance ('tol') from the i-th solution. After each step the ohmic constraint (see below) is checked and if not satisfied a new $\lambda_{o,o}$ is calculated and the procedure repeated.

The ohmic constraint consists in the following. Consider the standard Ohm's law for a magneto- fluid like:

$$\boldsymbol{E} = -\boldsymbol{v} \times \boldsymbol{B} + \eta \boldsymbol{J} + \nabla \phi$$

where **E** is the inductive electric field, **v** is the fluid velocity and $\phi$ the electrostatic potential.

Taking the parallel component to **B** of this equation and assuming helical symmetry and averaging over the helical flux surfaces in order to eliminate the electrostatic contribution it is found:

$$\lambda(\chi) = \frac{E_z <B_z>}{<\eta><B^2>} \qquad (4)$$

where < > denotes average over the helical flux surfaces, and in steady state with conservation of the magnetic flux the only remaining **E** component is a constant axial electric field.

Eq.(4) sets a strong constraint to the parallel current profile, $\lambda$, that should vanish where the axial magnetic field averaged over the helical surfaces also vanishes.

It is known from previous studies that the satisfaction of this constraint is unlikely for closed magnetic configurations [6]. In our perturbative approach (arrested at the first order) the main problem is that, whatever initial profile is chosen for $\lambda$, the zero of the axial magnetic field occurs radially well before the zero in the parallel current, and Eq.(4) could not be satisfied even in presence of a relatively large helical field (contributing to the averages). However, it should be reminded that we only use a first order perturbation expansion. Here we are mainly interested in studying the HGS equilibria and the ohmic constraint is considered as an "a posteriori" check.

The last comment to this section is that, although Eq.(2) allows the presence of a finite pressure, however the pressure effects have been shown to be negligible (see for example [7]) and we set the pressure to zero in this paper, so that we look in particular for force-free helical equilibria.

## 3. Numerical results

In this section we will first discuss the convergence issue of our method to the HGS states and then results of the numerical simulations for a specific set of parameters are given. We will concentrate to the case with an intermediate aspect ratio, A = 4 as in Ref.[1].

### 3.1 The non convergence of HGS states

As mentioned before the existence of SH RFP closed magnetic configurations in ohmic equilibrium has been proved to be problematic [6]. By numerically solving the problem (using the perturbative procedure outlined in section 2) it is found that the solution of the HGS equation does not lead to a converged state. The situation is illustrated in Fig.1.

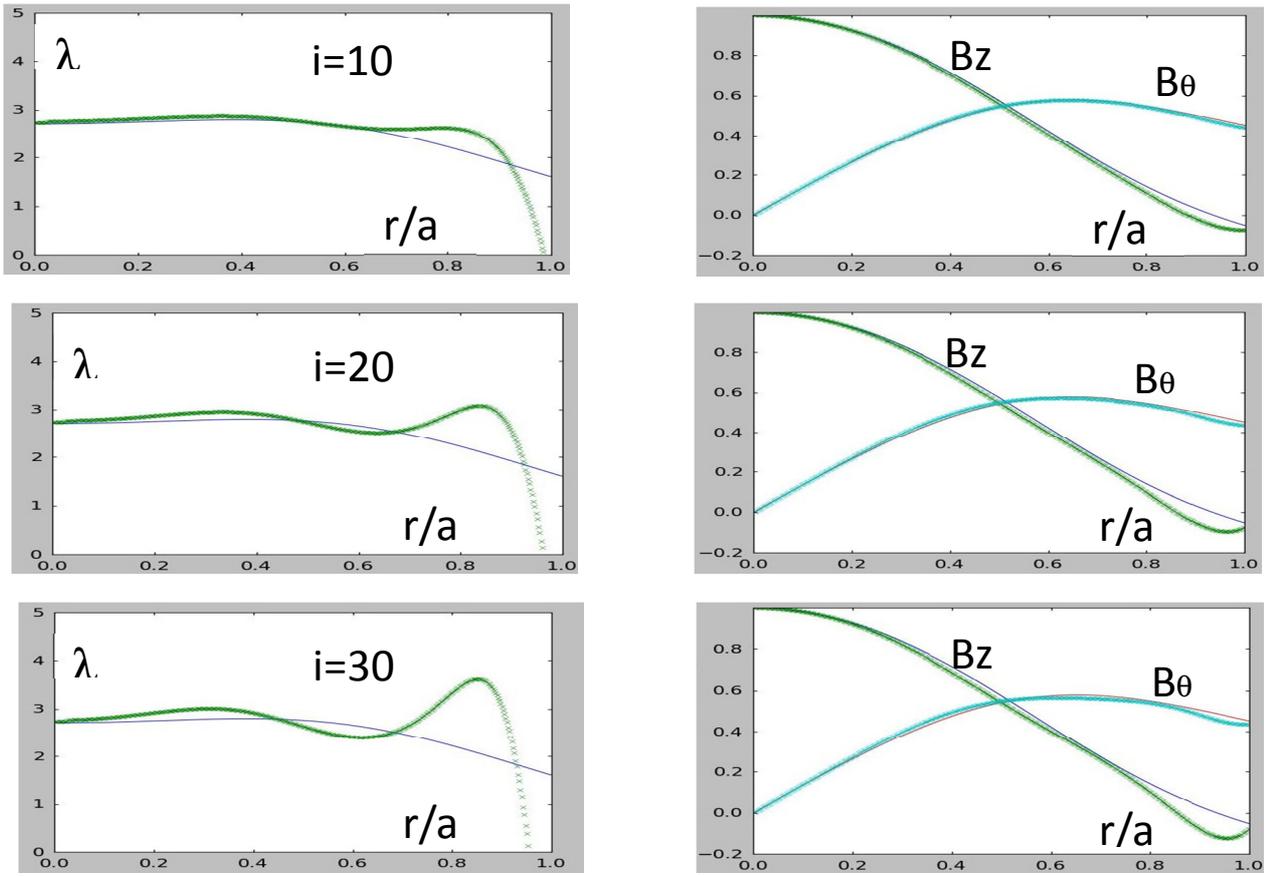

**Fig.1** λ and magnetic fields vs normalized radius after 10, 20 and 30 iterations of the HGS algorithm. The continuous lines correspond to the initial profiles.

The iteration of the HGS equation and of the Newcomb equation for the perturbation is inducing the reversal next to the wall of the initial (positive definite) λ profile. As mentioned, the ohmic constraint (Eq.(4)) is only checked a posteriori, after each iteration. However, it is never exactly satisfied. In fact although the zero of the λ profile is moving inward, the radial position of the zero of the toroidal field is also shifting toward the axis. The two radii ( the one where the axial field reverses and the one corresponding to the zero of the parallel current) are always different and Eq.(4), even considering the helical surface average, is not satisfied. Moreover the λ profile is developing a bump near the wall, whose amplitude increases with the iterations. This shows clearly the lack of convergence mentioned above. It is however interesting the change of sign of the parallel current profile near the edge.

This change is induced just by the form of the Newcomb perturbed solution and the quasi-neutrality condition and for example, does not require an enhanced dissipation at the edge to account for the current suppression in this region.

Also, the possible presence of a current bump just before the plasma boundary is a remarkable feature. Potentially this bump could be, for example, at the origin of a destabilization of the experimentally observed modes resonating between the reversal radius and the wall [8]. Since there is no true convergence we chose to stop our iterations when the radial field at the edge becomes negligible, as illustrated in Fig.2.

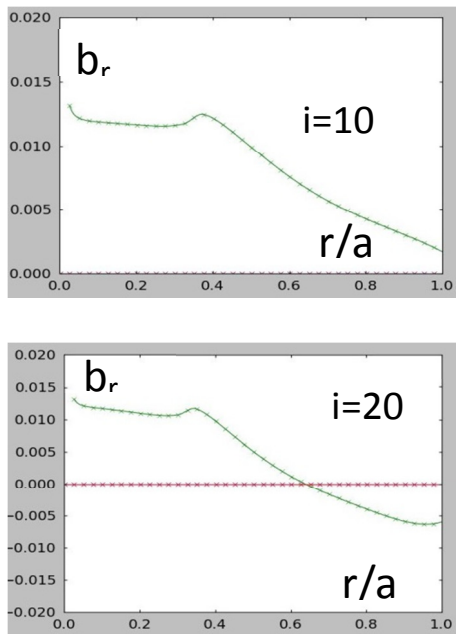

**Fig.2** Radial magnetic field vs. normalized radius for 10 and 20 iterations
 (the field profiles are those given in Fig.1 for the same number of iterations).

In the case shown in the figure this happens between 10 and 20 iterations. The configuration corresponding to a vanishingly small radial field at the wall can be considered marginally stable (i.e. a proper equilibrium solution), whereas if , for example, the zero is inside the wall the configuration is unstable to the given harmonic, according to the Newcomb's theorem [9].

Note that the marginal stable states, are reached after few iterations, corresponding to magnetic field profiles not very "far" from the initial relaxed states. In fact the initial SHR equilibrium is stable (the radial field has no zeros within the plasma boundary), as should be expected for a relaxed state obtained after an energy minimization procedure, however, the proposed HGS solver is finding the non axi-symmetric state that corresponds to a helical equilibrium satisfying the perfect conducting wall boundary conditions. This solution is therefore fully satisfying the requirements of the relaxation theory, allowing for the presence of well conserved quantities. It generally corresponds to a total magnetic energy content slightly above that of the initial axi-symmetric SHR state.

**3.2 From SHR to HGS states**

In this section the results of the solution of the coupled systems i.e. of single helical relaxation model (SHR), described in section 1 and of the helical Grad-Shafranov ( HGS) described in section 2 are presented in more detail. As already mentioned the parallel current profile ($\lambda_{0,0}$) obtained as SHR state is inserted as initial guess for the solution of the HGS.

The result of this exercise is shown in Fig.3 for a case with A=4 and a dominant mode n=7 (this mode corresponds to the internally resonant n=-7 reported for example in [10]).

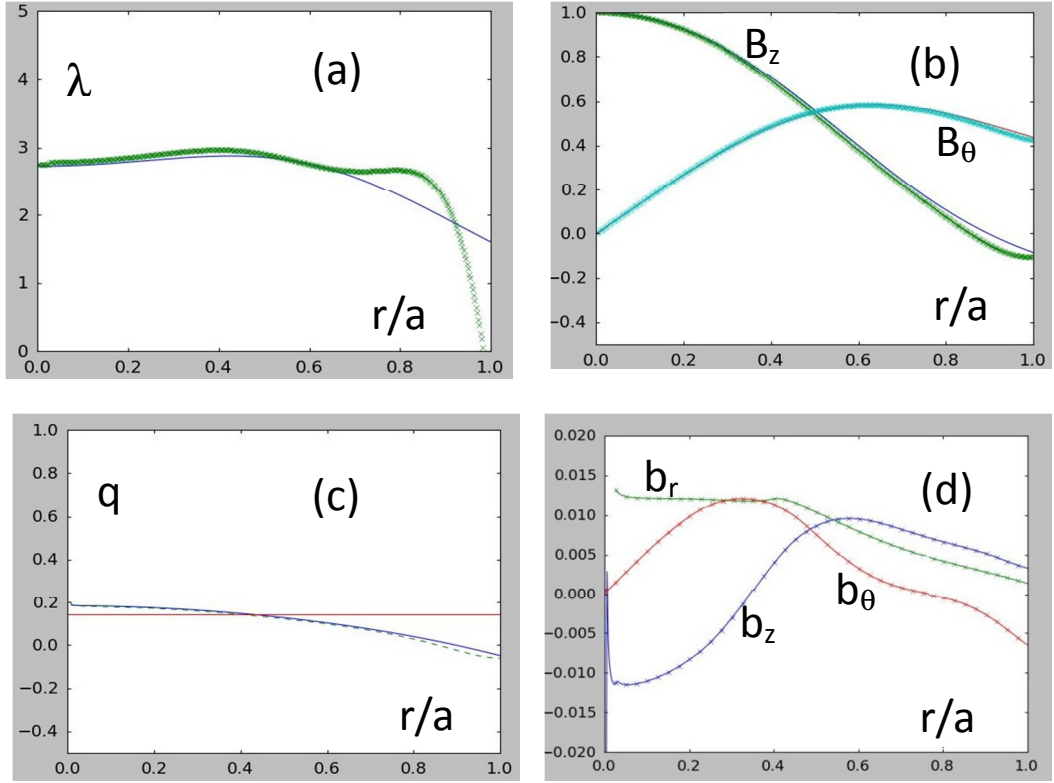

**Fig.3** Comparison between the input profiles for the magnetic field obtained within the SHR model and the output of the HGS solver: (a) $\lambda$ profiles, (b) magnetic fields, (c) q profile and (d) perturbed field components vs. normalized radius r/a (see text for more details).

In the frame labeled by (a) in figure 3 the initial $\lambda$ obtained by the SHR model (plain line) is compared with that obtained by solving the HGS (crosses). It is interesting to see that the ohmic constraint mainly affects the profile at the edge where, as we discussed in section 2, the parallel current should have a zero in correspondence of the zero of the toroidal magnetic field, following Eq.(4). In reality, as already mentioned, this constraint is not satisfied by the fields shown in Fig.3. As we anticipated it seems not possible to achieve an exact single helical ohmic equilibrium [6]. However it is interesting that our procedure leaves the $\lambda$ profile obtained from the relaxation theory almost unaffected for a large portion (more than 2/3) of the plasma minor radius. The field profiles (frame (b)) (again crosses are for the HGS solution) are not much affected and also the q profile (dashed line is the HGS solution) (frame (c)). As result also the macroscopic parameters of this configuration remain similar. This was a case with initial F=-0.27 and initial Θ=1.4 (see below for definitions) with d=2 and the SHR eigenvalue $\lambda_o$=0.8. After the changes introduced by the ohmic constraint the new values were: F=-0.36 and Θ=1.48, values that are almost aligned with the initial ones in the F-Θ plane (see Fig.5). The parameters F and Θ are defined as:

$$F = \frac{B_z(a)}{<B_z>} \quad and \quad \Theta = \frac{B_\theta(a)}{<B_z>}$$

with:

$$<B_z> = \frac{1}{\pi a^2} \int_0^a B_z \, r \, dr$$

and they describe in a simple way the macroscopic characteristics of the RFP plasmas, in particular the degree of axial field reversal, F, and the degree of pinch compression, Θ. The last frame (d) of Fig. 3 is showing the perturbed field components (having m=1 and n=7) obtained only within the HGS formalism. In Fig.4 it is seen that the λ profile (in red) (including the helical deformation) is not perfectly aligned with the constant χ surfaces (in blue) has it should be for an exact ohmic equilibrium, but we have already mention that this is not (and pretend not to be) a "perfect" ohmic helical state.

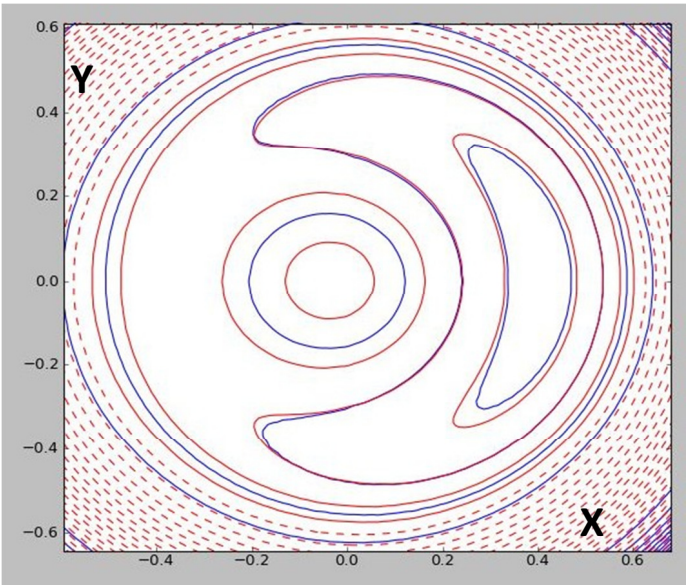

**Fig.4** Contours of constant χ and λ profiles in poloidal plane (normalized to minor radius, a, in cartesian (X,Y) coordinates).

From the figure the size and radial position of the 1/7 island can also be deduced. In this specific case the island width is about 30% of the plasma minor radius. Obviously this is determined by the free choice of the parameter ε as we discussed in section 2. This parameter was chosen here to produce islands of the same order of those measured in experiments [10]. In Fig.5 the F-Θ points corresponding to the initial relaxed states (circles) and the final quasi-ohmic states (squares) assuming A=4 are shown. The two sets are almost aligned and lie a little to the left of the experimental cloud (blue-gray region).

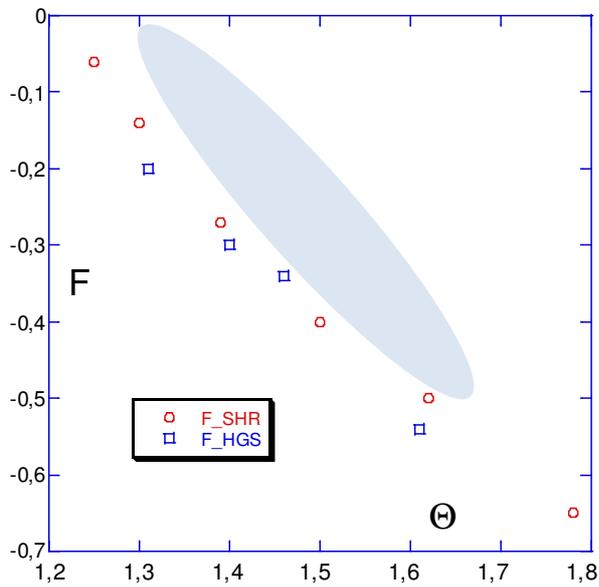

**Fig.5** F vs. Θ squares are final HGS and circles initial SHR states.
Also shown is the gray area of experimental data with A=4.3.

It can be seen that the HGS equilibria are, not surprisingly, aligned with the SHR states. It should also be noted that the experimental points lie to the right of the predicted states obtained with a dominant n=7 mode, especially at shallow F (F near 0) where instead, the n=7 is found dominant in experiments at this aspect ratio [10].

We remark also that the theoretical F-Θ points in Fig.5 would not change significantly by setting A=4.3.

However, the choice of the aspect ratio can instead affect the position of the resonance of the dominant mode, as the initial state, obtained by the relaxation model, is moved in the F-Θ plane. For example, by taking a dominant n=7 and by changing the eigenvalue for the two similar aspect ratios, the results are shown in Fig.6.

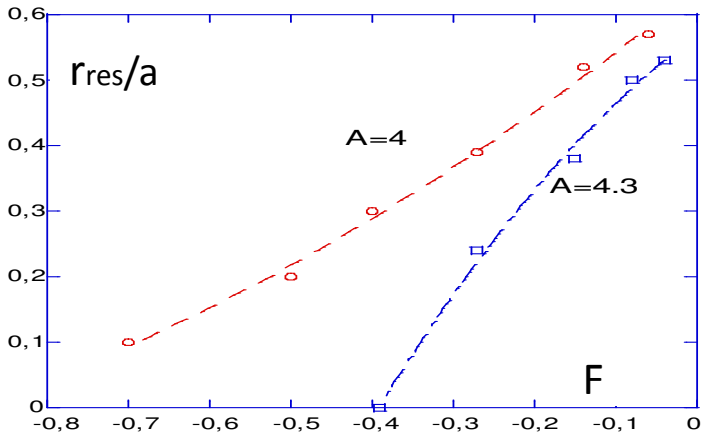

**Fig.6** Position of the normalized resonance radius of the n=7 (m=1) mode vs. F.

It is seen that the resonance moves toward the axis as F decreases (the value of q(0) is decreasing accordingly) till a point where the resonance of the mode disappears. At that point the n=8 becomes the most internal resonant mode. From the figure it can be seen that the F value at which the resonance disappears depends sensibly on the aspect ratio. In the figure the value A=4.3 corresponds to the RFX-mod device. The trend shown in Fig,6 can, in a simple way, explain two experimental observations: the first is that at shallow reversal (F vanishing) the most well developed SH states are observed, with the largest island structures in the plasma. This could be clearly connected with the fact that the resonance moves outwards and therefore the island could fill a larger plasma region (taking advantage from the low shear in the core). The second observation is the dependence on F of the dominant modes. At deep reversal higher n modes are observed in several devices [11,12,13]. Note therefore that the experimental cloud in Fig.5 is representative of SH states with different dominant helicities at the different F values. The results of Fig.6 could eventually be validated with the experiments, for example by comparison with tomographic data.

The fact that the size of the island depends on F (and therefore on the resonance position) is shown in Fig.7 for the HGS states at two F values.

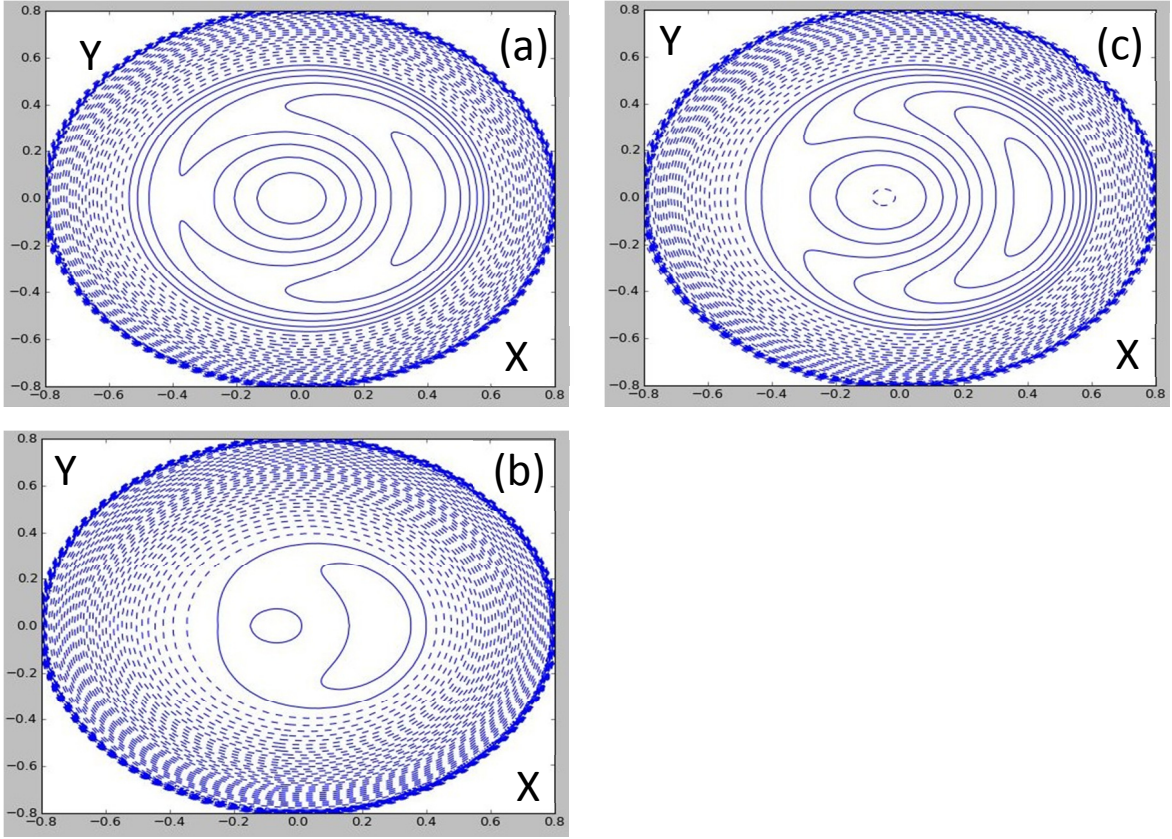

**Fig.7** HGS helical flux contours (cartesian coordinates as in Fig. 4) for F=-0.02 (a) and F=-0.21 (b) with A=4.3, n=7, ε=0.05 and (c) F=-0.02 with ε=0.1.

It can be seen that the case with the shallow F and with the resonance at a larger radius can support larger islands at higher perturbation amplitude (case(c) in the figure).

For ε=0.1 and F =-0.21 no HGS solution can be found since the resonance is too near to the axis and numerical problems arise.

### 3.3 Tuning of the SHR initial states to match experiments

In the previous section we assumed d=2 and the same profiles studied in [1].

Although this was shown, at intermediate aspect ratio, to be a satisfactory choice [1], by looking in more detail, and especially trying to find final HGS states that match well all the features seen in experiments, it is found that this choice should be revised.

In the attempt to do this fine-tuning we select the aspect ratio of the RFX-mod experiment [14], i.e. A=4.3.

The main experimental characteristics that we want to match are, more precisely:

( i ) the F-Θ curve (see Fig.5)

( ii ) the emergence of a dominant n=7 mode at very shallow reversal

(iii) the position of the resonance radius

All these characteristics depend quite sensibly on the choice of the exact value of the aspect ratio, as can be seen, for example, in Fig.6.

As seen in Fig,5 the F-Θ experimental curve is not satisfactorily matched starting from a SHR model with d=2, at shallow reversal. We consider therefore other possible choices. One particular simple choice that we found suitable is to set d=1 as in [2,3], however allowing some finite parallel current at the edge.

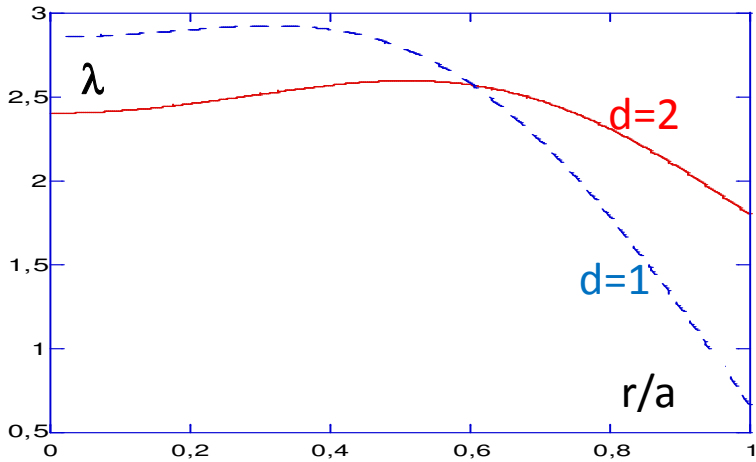

**Fig.8** SHR λ profile vs. normalized radius for d=2 and d=1 (with non zero parallel current at the edge).

The comparison between the two SHR profiles (with n=7 ) obtained with d=1 and d=2 and a non-zero current at the wall is shown in Fig.8. The case with d=1 corresponds to F=-0.04 and Θ=1.38. As can be seen from Fig.5 this match very well the experimental values at shallow F (grey region in the figure). Also the position of the resonance of the n=7 mode is, in this case, at around 0.3 of the normalized radius possibly in better agreement with the experimental results [15].

The initial SHR field profiles and those obtained solving the HGS are shown in Fig.9.

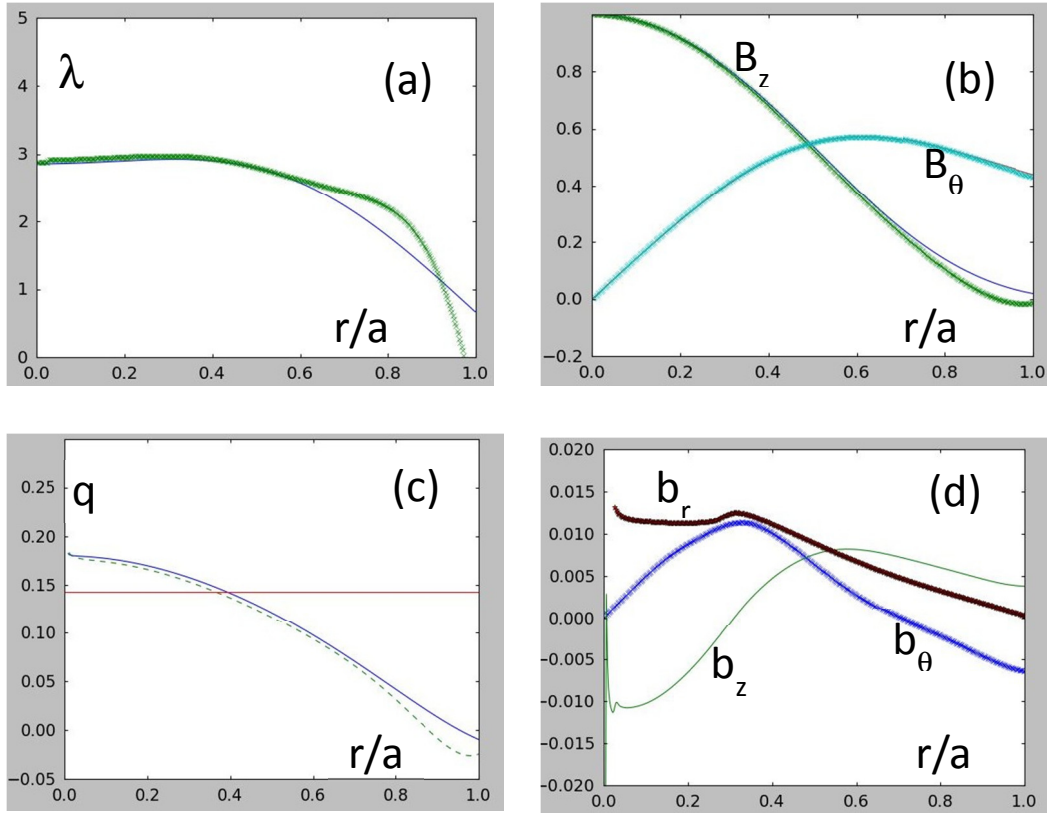

**Fig.9** Comparison between the input profiles for the magnetic field obtained within the SHR model and the output of the HGS solver: (a) $\lambda$ profiles, (b) magnetic fields, (c) q profile, (d) perturbed field components vs. normalized radius r/a (see text for more details).

In the frame (a) of the figure the initial $\lambda$ obtained by the SHR model (plain line) is compared with that obtained by solving the HGS (crosses). Similar comparisons are done in (b) and (c) for the magnetic field components and the q profiles, respectively. In (c) the resonance of the n=7 mode is indicated by the intersection of the horizontal (red) line with the q profile curve. Finally in (d) the perturbed magnetic field components (radial black, green poloidal and blue axial) are shown.

This choice of the SHR parameters also leads to the smallest deviation of the final HGS equilibrium from an ohmic state. In fact, as shown in Fig.10, the error, defined as the normalized radial distance between the surface average of the axial field crossing zero and the radial position of vanishing current (see Eq.(4)) is below 4%. It can also be seen that in the paramagnetic limit (F => 0) the ohmic constraint is almost exactly satisfied. In this case the matching is reached almost at the wall radius.

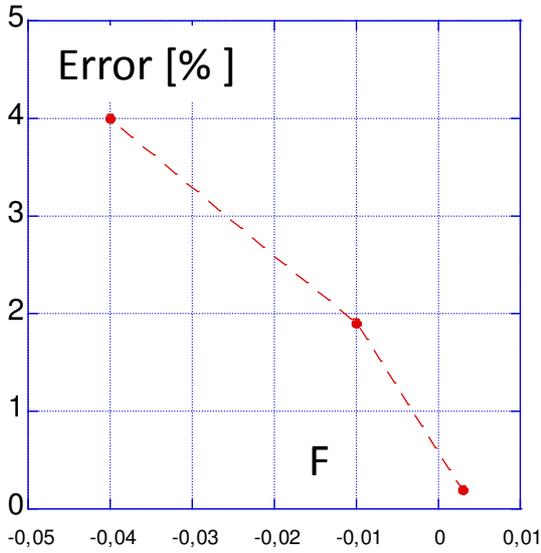

**Fig.10** Error (in %) (as defined in the text) vs. the F parameter.

However at F just below -0.05 the n=7 resonance is lost (see Fig.6 for a comparison with the case with d=2), in some disagreement with experimental data at this aspect ratio, where the n=7 mode is seen to be dominant till F around -0.2 [12]. Moving from the peaked $\lambda$ profile with d=1 to the more flat $\lambda$ with d=2 (see Fig.8) can accommodate this discrepancy. By commenting Fig.5, we already noted that, with our previous choice d=2, the F-$\Theta$ curve is satisfactorily matched for deeper F values.

Experimentally it is also observed that at deeper F the confinement degrades and so does the purity of the SH states. This is quite consistent with an enhanced turbulence and a flattening of the parallel current. Therefore our conclusion is that to match the experimental results, at this intermediate aspect ratio (but possibly in general), the d exponent should likely vary from 1 to 2 (see Fig.8). A unique choice for d, valid over the whole range of F values, seems in fact not able to satisfy all the critical aspects listed at the beginning of this section.

By looking at Eq.(1), this is, however really what should be expected applying the most general formulation of the relaxation theory. Our choice to consider separately the d=0,1 from the d=0,2 cases greatly simplify the solution of the problem, by allowing, for example, fully explicit boundary conditions and the determination of a unique eigenvalue, $\lambda_o$. Instead the general solution with d=0,1,2 would require to find all the 3 eigenvalues: $\lambda_o$, $\lambda_1$, $\lambda_2$, with at least 3 extra conditions to be imposed on the system and it would also require, in general, the solution of a differential problem with implicit boundary conditions. Therefore we decided to make the simpler assumption stated above.

It seems, however, that our simplifying assumption is not fully satisfactory for the comparison with the experimental data in the whole range of parameters. We leave a more detailed study of this subject for future work.

**Discussion and Conclusions:**

In this paper we look for the existence of ohmic helical equilibria in RFPs plasmas. The initial magnetic field profiles are obtained as result of a plasma relaxation process in which a single mode is dominant.

Within our modelling assumptions, we showed that the ohmic constraint is only approximately satisfied, independently of the initial condition, and that there is no convergence to a real ohmic steady state. This difficulty is due to the simple fact that the average toroidal field vanishes at a radial position shifted inward with respect to the zero of the $\lambda$ profile and also that this position could not be much influenced by the edge modulations of the parallel current.

It should be noted that (see Eq.(4)) the helical surface averages could in principle help in satisfying the

Ohmic constraint. However the helical radial field eigenfunction (see Fig.3 for example) is small from the axial field reversal to the wall and therefore it cannot give a significant contribution to the helical averages.

It has been recently proposed to actively excite the dominant mode by applying an external perturbation [15]. This could in principle help in increasing the dominant mode amplitude at the edge above the "natural" level. However there are problems associated with this technique. First of all, a large externally applied perturbation generates sidebands [16] that can be detrimental for the plasma confinement. Second, an increase of the edge radial field could lead to an enhanced plasma-wall interaction, again leading to a deterioration of the confinement and finally to a higher dissipation in the system.

Although the ohmic constraint is not "exactly" satisfied, we have found that the HGS solver, proposed here, acts on the parallel current profiles introducing peculiar features. In particular, it suppresses the current at the edge and also it tends to create a bump to compensate for this loss of current. These modifications are however not able to change significantly the F-Θ trajectory of the relaxed states, not surprisingly since both the SHR and the HGS are nearby stable force-free equilibria.

An interesting outcome of solving the HGS equation, by employing the proposed procedure, are the helical components of the field. Although the amplitude of the helical perturbation is prescribed (it is a free parameter in the model) the knowledge of the radial magnetic field profile allows to fully determine the helical flux function and therefore the island structure. Moreover, as we have shown, the model can predict the position of the island and its disappearance when the resonance is lost, for example by varying the F parameter. All these features could be compared in detail, as we start to do here, with the experimental results.

The fact that the dominant mode resonance tends to move outwards as the F parameter tends to zero, is consistent with the experimental observation that the largest islands are obtained for the n=7 mode, at shallow reversal [12]. Vice versa when F is decreasing (at deep reversal) the resonance of the n dominant mode is expelled from the plasma and the mode n+1 becomes the most internal resonant and therefore the next candidate to become the dominant helicity. This is also consistent with the experimental observation that the dominant n number depends on F and in particular, n increases as the toroidal field reversal becomes deeper and deeper ( see [11-13] ), although, for completeness, it should be noted that experimentally the purity and persistence of the SH states is much weaker at deeper reversal in comparison with the shallow reversal cases.

A careful examination of the A=4.3 case, corresponding to the RFX-mod device, shows, however, that it is not easy to match the experimental results at all F values assuming a unique value for the d exponent, determining the initial SHR state, in Eq.(1). It is likely instead (see section 3.3) that d should vary from 1 to 2 passing from shallow to deep reversal. This could be also reasonably linked to the tendency of the parallel current profile to be flatter at deep F due to the increase of the turbulence level in experiment.

It should also be remarked that the fact that the best experimental energy confinement is found for the shallow reversed cases [17] with a n=7 dominant mode, is consistent with the model prediction that the volume of the island is maximized and larger amplitudes of the dominant modes are allowed.. The increase in the amplitude of the dominant mode can bring to a full reconnection of the core with the so called "separatrix expulsion", with associated further increase of the island volume and also with a possible direct effect on the stochastic transport due to a chaos healing effect [18]. In a more recent paper the transport improvement was mainly associated with the reduction of the amplitude of the secondary modes, but in any case a clear correlation of the plasma energy content with the amplitude of the dominant mode was strengthened [2].

A final but important remark should be done about the existence of exact HGS ohmic states.

The results presented here, showing the non convergence to such states (at least in a mathematical/numerical sense), cannot be considered as a rigorous proof of their non-existence. In fact, first of all, in this paper only a first order perturbation theory is assumed whereas, as discussed for example in [19], at least a second order expansion is needed. Second, in 3D numerical simulations, ohmic single helicity visco-resistive states are shown to exist [15,19]. Therefore the present model is incomplete under this respect and cannot give a uncontroversial proof of existence (or non-existence) of the single helicity ohmic states. However, as we have shown here, the model is rich enough to account for many important features and trends of the single helical states as detected in experiments and can help in the interpretation of several aspects, even if within our numerical procedure only "quasi-ohmic" states are found, possibly with the single exception of the paramagnetic pinch (see section 3.3) with vanishing toroidal field at the wall.